\documentclass[reprint,nofootinbib,superscriptaddress,amsmath,amssymb,aps]{revtex4-1}
\usepackage{graphicx}
\usepackage{rotating}
\usepackage{dcolumn}
\usepackage{bm}
\usepackage{color}
\usepackage{mathptmx, textcomp}
\usepackage[latin1]{inputenc}
\usepackage{braket}
\usepackage[colorlinks,linkcolor=magenta,anchorcolor=cyan,citecolor=blue]{hyperref}
\usepackage[multiple]{footmisc}

\def\be{\begin{equation}}
\def\ee{\end{equation}}
\def\ba{\begin{eqnarray}}
\def\ea{\end{eqnarray}}

\def\nn{\nonumber}
\def\lf{\left}
\def\rt{\right}

\newcommand{\eq}[1]{(\ref{#1})}

\def\dbar{{\mathchar '26\mkern -10mu\delta}}

\def\lf{\left}\def\rt{\right}\def\q{\theta} \def\w{\omega}\def\r {\rho}     \def\p {\pi}   \def\d {\delta} \def\f {\phi}  \def\h {\eta} \def\j {\varphi} \def\k {\kappa} \def\l {\lambda} \def\z {\zeta} \def\x {\xi} \def\c {\chi}   \def\m {\mu} \def\pd {\partial}\def\p {\pi} \def \inf {\infty}  \def \e { \varepsilon}
\def\Q{\Theta} \def\W{\Omega} \def\Y {\Psi}    \def\S {\Sigma}  \def\F {\Phi}      \def\grad{\nabla}\def\.{\cdot}
\def\math {\mathcal}
\begin{document}

\title{Constraints on High-Order Gravitational Theories from Weak Cosmic Censorship}
\author{Jie Jiang}
\email{jiejiang@mail.bnu.edu.cn}
\affiliation{Department of Physics, Beijing Normal University, Beijing 100875, China\label{addr2}}
\author{Aofei Sang}
\email{202021140021@mail.bnu.edu.cn}
\affiliation{Department of Physics, Beijing Normal University, Beijing 100875, China\label{addr2}}
\author{Ming Zhang}
\email{Corresponding author. mingzhang@jxnu.edu.cn}
\affiliation{Department of Physics, Jiangxi Normal University, Nanchang 330022, China}
\date{\today}

\begin{abstract}
Weak cosmic censorship conjecture (WCCC) is a basic principle that guarantees the predictability of spacetime and should be valid in any classical theories. One critical scientific question is whether the WCCC can serve as a constraint to the gravitational theories. To explore this question, we perform the first-order Sorce-Wald's gedanken experiments to test the WCCC in the higher-order gravitational theories and find that there exists a destruction condition $S_\text{ext}'(r_h)<0$ for the extremal black holes. To show the power of this condition, we evaluate the constraints given by WCCC in the quadratic and cubic gravitational theories. Our investigation makes an essential step toward applying WCCC to constrain the modified gravitational theories, and opens a new avenue to judge which theory is reasonable.
\end{abstract}
\maketitle
\textit{Introduction.}--- The weak cosmic censorship conjecture (WCCC) \cite{Penrose:1969pc} proposed by Penrose is a basic principle to keep the predictability of the spacetimes. It states that the spacetime singularity should be hidden inside the event horizon of black holes. To examine this conjecture, known as the old version of the gedanken experiments proposed by Wald \cite{Wald:1974wl}, is to throw charged or spinning particle into a black hole to destroy the horizon. Calculations for the extremal Kerr-Newman black hole proved that this experiment respects the WCCC. After that, many investigations have tried to challenge this conjecture in various ways. Recently, Sorce and Wald \cite{SW} proposed a new gedanken experiment to test WCCC by a general matter perturbation, where the full dynamical process of matter and gravity was considered. Given the null energy condition (NEC) of the matters, they showed that the WCCC is valid for the nearly extremal Kerr-Newman black hole under the second-order approximation of perturbation.

Besides general relativity, several high-order theories are also proposed for the investigations of gravity. In the low-energy effective field theory (EFT), the Einstein-Maxwell theory should be modified by adding higher-order derivative terms as quantum corrections in the regime of higher energy. In cosmology and astronomy, high-order modified gravitational theories, such as Weyl gravity \cite{Mannheim:2010ti}, $f(R)$ gravity \cite{TP} and modified Gauss-Bonnet gravity \cite{Nojiri:2005jg}, are often introduced to explain the observational evidence associated with early-time inflation, late-time acceleration, etc. However, there is very little understanding of which gravitational theories being reasonable. At the quantum level, these theories can be effectively constrained by the necessity of ghost-free \cite{Hawking:2001yt} or weak gravity conjecture \cite{Vafa1, Vafa2}. At the classical level, to guarantee the predictability of spacetime, it is natural to demand that any admitted theories satisfy the WCCC. One critical scientific question is whether the WCCC can give a constraint to the high-order theory. Recently, an attempt to this question was given in Ref. \cite{Chenprl}. Using Sorce-Wald's method under the first-order approximation, the authors showed that the WCCC fails for some possible generations; thus they argued that the WCCC provides a constraint to the high-order EFT. Straightforwardly, this result is indeed remarkable. However, in Ref. \cite{Jiang:2021ohh}, we reexamined their letter and found that key errors occur. After correcting the errors, their discussion cannot give the constraint to the high-order theories. In this letter, we will extend the discussion to a more general higher-order gravitational theory and show which condition should be imposed by demanding the validity of WCCC.

\textit{Testing WCCC in the higher-order gravitational theory.}---In the following, we consider a $n$-dimensional gravitational theory coupled to an electromagnetic field $\bm{A}$ with the Lagrangian $n$-form $\bm{L}=\bm{\epsilon}\math{L}(g_{ab}, R_{abcd}, F_{ab})$ with $\bm{F}=d\bm{A}$, in which $\math{L}$ is an analytical function of the contractions of $R_{abcd}$ and $F_{ab}$. Denoting $\f=\{g_{ab}, \bm{A}\}$, the variation of $\bm{L}$ gives
\ba\begin{aligned}
\d \bm{L}=\bm{E}_\f \d\f+d\bm{\Q}(\f, \d\f)\,,
\end{aligned}\ea
where $\bm{E}_\f\d\f=-\bm{\epsilon}\left[(1/2)T^{ab}\d g_{ab}+j^a\d A_a\right]$ with the stress-energy tensor
\ba\begin{aligned}
T^{ab}= 2 E_R^{cde(a}R_{cde}{}^{b)}+4\grad_c\grad_d E_R^{(a|c|b)d}-2E_F^{c(a}F^{b)}{}_c-g^{ab}\math{L}
\end{aligned}\nn\\\ea
and electric current $j^b=2\grad_a E_F^{ab}$ for the extra matter source, and $\bm{\Q}(\f, \d\f)$ is the symplectic potential $(n-1)$-form locally constructed out of $\f, \d \f$ and their derivatives. Here we have denoted
\ba
E_R^{abcd}=\frac{\pd\math{L}}{\pd R_{abcd}}\,,\quad E_F^{ab}=\frac{\pd\math{L}}{\pd F_{ab}}\,.
\ea
For the source-free solution, we have $T_{ab}=0$, $j^b=0$.  The Noether current $\bm{J}_\z$ associated with $\z^a$ is defined by
\ba\begin{aligned}\label{J1}
\bm{J}_\z\equiv\bm{\Q}(\f, \math{L}_\z\f)-\z\.\bm{L}\,.
\end{aligned}\ea
With a tedious but straightforward calculation, it can also be written as (See Sec. I of Supplemental Material for full expressions)
\ba\begin{aligned}\label{J2}
\bm{J}_\z=\bm{C}_\z+d\bm{Q}_\z
\end{aligned}\ea
with $ (\bm{C}_\z)_{a_2\cdots a_n}=\bm{\epsilon}_{a a_2\cdots a_n}(T_b{}^a+A_b j^a)\z^b\,.$

Next, we consider an asymptotically flat axisymmetric stationary black hole solution which contains a bifurcate Killing horizon $H$ and satisfies the source-free equation of motion (EOM) $T_{ab}=j^b=0$. The Killing vector field of horizon is given by
\ba\begin{aligned}
\x^a=t^a+\W_H^{(\m)} \j^a_{(\m)}\,,
\end{aligned}\ea
where $t^a$ is a stationary Killing field related to an asymptotic time translation, $\j_{(\m)}^a$ is an axial Killing field, and $\W_H^{(\m)}$ is the angular velocity of the horizon.

For these black hole solutions, the mass $M$, angular momentum $J_{(\m)}$, electric charge $Q$, and entropy $S$ are defined by \cite{IW}
\ba\begin{aligned}\label{MJ}
M&=\int_\infty \left(\bm{Q}_t-t\.\bm{B}\right)\,,\quad J_{(\m)}=-\int_\inf  \bm{Q}_{\j_{(\m)}}\,,\\
Q&=-\int_{\inf} \bm{\epsilon}_{aba_3\cdots a_n} E_F^{ab}\,,
S=-2\p\int_\math{B}\bm{\epsilon}_{aba_3\cdots a_n} \bm{\hat{\epsilon}}_{cd} E_R^{abcd}\,,
\end{aligned}\ea
in which $\hat{\bm{\epsilon}}$ is the binormal of the bifurcate surface $\math{B}$, and $\bm{B}$ is a $(n-2)$-form defined by
\ba\begin{aligned}
\d\int_\inf t\.\bm{B}=\int_{\inf}t\.\bm{\Q}(\f, \d\f)\,.
\end{aligned}\ea
The existence of $\bm{B}(\f)$ for an asymptotically flat spacetime in $\math{L}(g_{ab}, R_{abcd}, F_{ab})$ gravity has been proved in Sec. II of Supplemental Material. Moreover, with a calculation similar to Ref. \cite{IW} (See Sec. III of Supplemental Material), we can show that these black holes obey the first law
\ba\begin{aligned}\label{firstlaw}
\d M&=T \d S+\W_H^{(\m)}\d J_{(\m)}+\F_H \d Q\\
&=T \d S+\Y_i \d \c^i\,,
\end{aligned}\ea
in which we denoted $\c^i=\{J_{(\m)}, Q\}\,,$ $\Y_i=\{\W^{(\m)}_H, \F_H\}$, $\F_H\equiv-\left.A_a\x^a\right|_H$ is the electric potential of the black holes, and $T=\k_H/2\p$ with the surface gravity $\k_H$ on horizon is the temperature of the black holes.

For a stationary black hole, there exists a blackening factor $f(r, M, \c)$ such that $f(r, M, \c)$ is positive outside the horizon and vanishes on the horizon, i.e., $f(r_h, M, \c)=0$ and $\pd_rf(r_h, M, \c)\geq 0$. From the variation of the entropy $S=S(r_h,M,\c)$ and $f(r_h, M, \c)=0$, together with the first law \eq{firstlaw}, we can obtain the relationships
\ba\begin{aligned}\label{TY}
T&=\frac{\pd_r f_H}{\pd_M S_H \pd_r f_H-\pd_M f_H \pd_r S_H}\,,\\
\Y_i&=\frac{\pd_i S_H \pd_r f_H-\pd_i f_H \pd_r S_H}{\pd_M f_H \pd_r S_H-\pd_M S_H \pd_r f_H}\,,
\end{aligned}\ea
in which the subscript $H$ denotes the quantities evaluated at $r=r_h$. The function $S(r_1, M, \c)\equiv-2\p\int_{S_1}\bm{\epsilon}_{aba_3\cdots a_n} \bm{\hat{\epsilon}}_{cd} E_R^{abcd}$ for any codimension-2 surface $S_1$ with $r=r_1$, and it becomes the black hole entropy iff $r_1=r_h$.

Now, we consider a situation where the extremal stationary black hole described by $f(r, M, \c)$ is perturbed by some accreting process and settles down to another stationary state with different $M'$ and $\c'$ at sufficiently late times (i.e., the spacetime satisfies the stability condition \cite{SW}). With a similar consideration of Ref. \cite{SW}, we assume that all of the matter goes into the black hole. Then, we consider a one-parameter family labeled by $\l$ in which every configuration $\f(\l)$ describes the above accreting process. At late times, the geometry of configuration $\f(\l)$ is described by a stationary state with $f(r, M(\l), \c(\l))$. Therefore, the key step to test WCCC is checking the sign of the minimal value
\ba\begin{aligned}
h(\l)\equiv f(r_m(\l), M(\l), \c(\l))
\end{aligned}\ea
of the function $f(r, M(\l), \c(\l))$. Here $r_m(\l)$ is the minimal radius satisfying $\pd_r f(r_m(\l), M(\l), \c(\l))=0.$
Then, we have
\ba\begin{aligned}\label{hl}
&h(\l)=f_m+\l \pd_Mf_m \left(\d M+\frac{\pd_i f_m}{\pd_Mf_m}\d \c^i\right)+\math{O}(\l^2)\,,
\end{aligned}\ea
where the subscript $m$ denotes the quantity evaluated at $r=r_m$. When the background spacetime is an extremal black hole, we have $r_m=r_h$, and therefore $f_m=f_H=0$. Using the expression of $\Y_i$ in Eq. \eq{TY}, we can obtain
\ba\begin{aligned}\label{hlf}
h(\l)=\l \pd_Mf_H \lf[\d M-\Y_i^\text{ext}\d \c^i\right]+\math{O}(\l^2).
\end{aligned}\ea

Next, we perform Sorce-Wald's method \cite{SW} to derive the first-order perturbation inequality. With a same setup, we choose a hypersurface $\S=\math{H}\cup \S_1$, in which $\math{H}$ is a portion of the future horizon connecting an early cross-section $B_0$ where the perturbation vanishes and another cross-section $B_1$ at sufficiently late times; $\S_1$ is a late-time spacelike surface connecting $B_1$ and infinity. Considering the stability assumption, the geometry on $\S_1$ is described by the stationary state with $f(r, M(\l), \c(\l))$ and therefore we have $\bm{E}_{\f}(\l)=0$ on the hypersurface $\S_1$.

From the two expressions of $\bm{J}_\z$ in Eqs. \eq{J1} and \eq{J2}, it is easy to get \cite{SW}
\ba\begin{aligned}\label{varid}
&d[\d \bm{Q}_\z-\z\.\bm{\Q}(\f,\d \f)]\\
&=\bm{\w}(\f, \d \f, \math{L}_\z \f)-\x\. \bm{E}_\f\d \f-\d\bm{C}_\z
\end{aligned}\ea
with the symplectic current $(n-1)$-form
$\bm{\w}(\f, \d_1\f, \d_2\f)=\d_1\bm{\Q}(\f, \d_2\f)-\d_2\bm{\Q}(\f, \d_1\f)\,.$ After replacing $\z^a$ by $\x^a$, integration of Eq. \eq{varid} on $\S$ gives
\ba\begin{aligned}
\d M-\W_H^{(\m)} \d J_{(\m)}=-\int_\math{H}\bm{\epsilon}_{ea_2\cdots a_n}\lf(\d T_a{}^e+A_a\d j^e\rt)\x^a\,,
\end{aligned}\ea
where we used the setup that $\d\f=0$ on $B_0$ and the fact that the background spacetime is source-free and stationary, i.e., $T_{ab}=j^b=\bm{E}_\f=0$ and $\math{L}_{\x}\f=0$. Since $\x^a A_a$ is constant on $\math{H}$, we can pull it out of the integral. The integral $\d Q_\text{flux}=\int_\math{H}\d (\bm{\epsilon}_{ebcd}j^e)$ is just the total flux of electric charge through $\math{H}$. Using the EOM $j^b(\l)=2\grad_aE_F^{ab}(\l)$, it is not hard to verify that $\d Q_\text{flux}=\d Q$. Then, we have
\ba\label{var1eq1}\begin{aligned}
\d M-\Y_i \d \c^i&=-\int_\math{H}\bm{\epsilon}_{ea_2\cdots a_n}\d T_a{}^e\x^a=\int_\math{H}\bm{\tilde{\epsilon}} \d T_{ab}k^a \x^b\,,
\end{aligned}\ea
where $k^a \propto \x^a$ is a future-directed normal vector on $\math{H}$, $\tilde{\bm{\epsilon}}$ defined by $\bm{\epsilon}_{a_1 a_2\cdots a_n}\equiv-nk_{[a_1}\tilde{\bm{\epsilon}}_{a_2\cdots a_n]}$ is the induced volume element on $\math{H}$. According to the NEC $\d T_{ab}k^a k^b\geq 0$, Eq. \eq{var1eq1} yields
\ba\label{ineq1}
\d M-\Y_i \d \c^i\geq 0\,.
\ea
Considering that the background black hole is extreme, we have $\Y_i=\Y_i^\text{ext}$. From Eq. \eq{hlf}, we can see that if $\pd_Mf_H^\text{ext}\leq0$, the above inequality implies $h(\l)\leq0$ under the first-order approximation of $\l$ and thus the black hole cannot be destroyed. However, if $\pd_Mf_H^\text{ext}>0$, we have $h(\l)\geq0$, and therefore the extremal black hole can be destroyed.

Since $M$ and $\c$ are defined at infinity, calculation of $\pd_M f_H^\text{ext}$ needs the connection between the horizon and infinity, which implies that we need the entire solution of the spacetimes. However, for most modified gravity, it is very difficult to get an analytical solution. To overcome this difficulty, we need to convert the destruction condition $\pd_M f_H^\text{ext}>0$ into the one that only depends on the event horizon, and therefore we can evaluate it from the near-horizon solution.

To do this, we consider a nearly extremal black hole where $\e=\pd_r f_H$ is a small positive quantity. From Eq. \eq{TY}, we have
\ba\begin{aligned}
T\simeq-\frac{\e}{\pd_M f_H^\text{ext}\pd_r S_H^\text{ext}}+\math{O}(\e)\,.
\end{aligned}\ea
Considering that $T>0$ for any non-extremal black hole, we must have $\pd_M f_H^\text{ext} \pd_r S_H^\text{ext}<0$. For instance, for static black hole with the line element
\ba\begin{aligned}
ds^2=-f(r)dv^2+2\m(r) dv dr+r^2(d\q^2+\sin^2\q d\f^2)\,,
\end{aligned}\ea
$\pd_M f_H^\text{ext} \pd_r S_H^\text{ext}\propto -\m^\text{ext}(r_h)$ is a negative quantity. Then, the destruction condition $\pd_Mf_H^\text{ext}>0$ is equivalent to $S'_\text{ext}(r_h)=\pd_r S_H^\text{ext}<0$. Different to $\pd_Mf_H^\text{ext}$, the quantity $S'_\text{ext}(r_h)$ only depends on the near-horizon solution of the extremal black holes. Therefore, the destruction condition $S_\text{ext}'(r_h)<0$ can be analytically discussed in the concrete theory.

It is worth noting that the destruction of the black holes is not equivalent to a violation of WCCC. To check the violation of WCCC, we also need to show whether there exists a spacetime singularity in the late-time geometry under the destruction case. For the theory that contains a central singularity, the destruction condition $S_\text{ext}'(r_h)<0$ is also the violation condition of WCCC. If we take the WCCC as a universal physical principle, then any black hole solution in the admitted theories should satisfy it.

To see the constraint from the WCCC, we consider the following two modified gravitational theories:\\
\textbf{Quadratic gravity:} Consider a four-dimensional quadratic gravity coupled to the electrodynamics with the Lagrangian
\ba\begin{aligned}
\math{L}=\frac{1}{2} R+c_1 R^2+c_2 R_{ab}R^{ab}+c_3 R_{abcd}R^{abcd}-\frac{1}{4}F_{ab}F^{ab}\,.\quad\quad
\end{aligned}\ea
After imposing the extremal condition $f(r_h)=f'(r_h)=0$ and $\m(r_h)=1$, the near-horizon solution of the extremal black holes can be obtained and it is given by (See Sec. IV A of Supplemental Material for details)
\ba\begin{aligned}
r_h&=\frac{q}{\sqrt{2}}\,,\quad\quad f''(r_h)=\frac{2}{r_h^2}\,,\\
f^{(3)}(r_h)&=-\frac{12(16c_2+64c_3+r_h^2)}{r_h^3(12c_2+48c_3+r_h^2)}\,,\\
\m'(r_h)&=-\frac{8(c_2+4c_3)}{r_h(11c_2+48c_3+r_h^2)}\,,
\end{aligned}\ea
in which we have set $q\geq0$ without loss of generality. Then, it is not hard to obtain
\ba\begin{aligned}\label{SEXP1}
S'_\text{ext}(r_h)=\frac{16\p^2}{r_h}[r_h^2+4(c_2+4c_3)]\,.
\end{aligned}\ea
By analyzing the asymptotic properties of EOM at $r=0$ (See Sec. IV B of Supplemental Material), we can see that there must be a center singularity in the spherically symmetric charged spacetime for the quadratic gravity, which implies that the violation condition of WCCC is given by $S'_\text{ext}(r_h)<0$.

From Eq. \eq{SEXP1}, we can see that if $c_2+4c_3<0$, the extremal black holes with $r_h^2<-4(c_2+4c_3)$ will violate the WCCC. Therefore, the validity of WCCC can give the parameter constraint
\ba\begin{aligned}
c_2+4c_3\geq 0\,.
\end{aligned}\ea
Note that our result is invariant under the transformation $c_2\to c_2-4c, c_3\to c_3+c$ with $c$ a constant, which reflects the fact that Gauss-Bonnet term $\math{L}_\text{GB}=R^2-4R_{ab}R^{ab}+R_{abcd}R^{abcd}$ in four-dimensional spacetime is only a topological term and does not contribute to the EOM.\\
\textbf{Cubic gravity:} Next, we consider a four-dimensional cubic gravity coupled to the electrodynamics with the Lagrangian
\ba\begin{aligned}
\math{L}=\frac{1}{2} R+c_1 R_{ab}^{cd}R_{cd}^{ef}R_{ef}^{ab}+c_2 R R_{ab}^{cd}R_{cd}^{ab}-\frac{1}{4}F_{ab}F^{ab}\,.
\end{aligned}\ea
With a straightforward calculation, the near-horizon solution of the extremal black hole is given by
\ba\begin{aligned}\label{sol2}
q^2&=\frac{2(r_h^4+24c_1+16c_2)}{r_h^2}\,, f''(r_h)=\frac{2}{r_h^2}\,,\cdots\,.
\end{aligned}\ea
The full expressions is given in Sec. V B of Supplemental Material. Then, we have
\ba\begin{aligned}
S'_\text{ext}(r_h)=\frac{8\p^2}{r_h^3}\frac{r_h^8-96 c_1 r_h^4+192 \left(9 c_1^2+28 c_2 c_1+20 c_2^2\right)}{24 \left(c_1+2 c_2\right)+r_h^4}.\quad\quad
\end{aligned}\ea
We can prove that there also exists a central singularity in the charged cases for this theory. Then, by demanding $S_\text{ext}'(r_h)\geq0$ for any extremal black holes in the admitted theory, it is not hard to obtain the parameter constraint as
\ba\begin{aligned}
c_2\geq0\quad \text{and}\quad c_1\leq 10c_2\,.
\end{aligned}\ea
(See details in Sec. V of Supplemental Material.)

\textit{Conclusion.}---As a universal physical principle, it is a critical scientific question whether the WCCC can serve as a constraint to the gravitational theories. To explore this question, B. Chen $et.\,al$ examined the WCCC in the low-energy EFT and found that there is a parameter bound by demanding the validity of WCCC \cite{Chenprl}. However, in Ref. \cite{Jiang:2021ohh}, we found there exists an error in their letter. After correcting it, their discussion cannot give the constraint to the gravitational theory.

In this letter, we extended the discussion into a more general higher-order gravitational theory. As a result, we found that the extremal charged black holes satisfying the condition $S'_\text{ext}(r_h)<0$ can be destroyed in the first-order Sorce-Wald's gedanken experiments. As the applications of this destruction condition, we performed it into the quadratic and cubic gravities and got the constraint of the theories by demanding the validity of the WCCC. Our result provides evidence that the WCCC can be used to constrain the high-order gravitational theories, and also offers a feasible method to calculate the constraint.

Unlike Ref. \cite{Chenprl}, the main object of our letter is the modified theory with finite higher-order derivative corrections. From this perspective, our investigation opens a new avenue to judge which modified theory in cosmology and astronomy is reasonable.
\\
\\
\\
We acknowledge financial supports from the National Natural Science Foundation of China (Grants No. 11775022, 11873044 and 12005080).

\onecolumngrid
\newpage
{\textbf{\LARGE{Supplemental Material:}}}

\section{Explicit forms of EOM, $\Q$, $Q_\z$ and $C_\z$ for the $\math{L}(g_{ab}, R_{abcd}, F_{ab})$ gravity}\label{secB}

The Lagrangian $n$-form is given by
\ba\begin{aligned}
\bm{L}=\bm{\epsilon}\math{L}(g_{ab}, R_{abcd}, F_{ab})
\end{aligned}\ea
with $\bm{F}=d\bm{A}$. The variation of $\bm{L}$ gives
\ba\begin{aligned}
\d \bm{L}=\bm{E}_\f \d\f+d\bm{\Q}(\f, \d\f)\,.
\end{aligned}\ea
On the other hand, we have
\ba\begin{aligned}\label{dL}
\d\bm{L}&=\bm{\epsilon}\d\math{L}+(\d\bm{\epsilon})\math{L}\\
&=\bm{\epsilon}\d\math{L}+\frac{1}{2}\bm{L}g^{ab}\d g_{ab}\,.
\end{aligned}\ea
For the first term of Eq. \eq{dL}, we have
\ba\begin{aligned}
\d\math{L}&=E_R^{abcd}\d R_{abcd}+E_F^{ab}\d F_{ab}+A^{ab}\d g_{ab}\\
&=E_R^{abcd}R_{abc}{}^e\d g_{de}-2E_R^{acbd}\grad_d\grad_c\d g_{ab}+2E_F^{ab}\grad_a\d A_b+A^{ab}\d g_{ab}\\
&=(E_R^{cdea}R_{cde}{}^b-2\grad_c\grad_dE_R^{acbd}+A^{ab})\d g_{ab}-2\grad_aE_F^{ab}\d A_b+\grad_d v^d,
\end{aligned}\ea
in which
\ba\begin{aligned}
A^{ab}&=\frac{\pd \math{L}}{\pd g_{ab}}\,,\\
\dbar v^d&=2\grad_c E_R^{adbc}\d g_{ab}-2 E_R^{acbd}\grad_c \d g_{ab}+2 E_F^{db}\d A_b.
\end{aligned}\ea
Using the relationship
\ba\begin{aligned}
\grad_d \dbar v^d=d\star \bm{\dbar v}\,,
\end{aligned}\ea
we can obtain
\ba\begin{aligned}\label{ETheta}
\bm{\Q}(\f, \d\f)&=\bm{\Q}^\text{grav}(\f, \d g)+\bm{\Q}^\text{e.m.}(\f,\d\bm{A})\,,
\end{aligned}\ea
with
\ba\begin{aligned}
\bm{\Q}^\text{grav}_{a_2\cdots a_n}(\f, \d g)&=2\bm{\epsilon}_{ca_2\cdots a_n}\left(E_R^{abcd}\grad_b\d g_{ad}+\d g_{bd} \grad_a E_R^{abcd}\right)\,,\\
\bm{\Q}^\text{e.m.}_{a_2\cdots a_n}(\f, \d \bm{A})&=2\bm{\epsilon}_{aa_2\cdots a_n}E_F^{ab}\d A_b\,.\\
\end{aligned}\ea
Using the results
\ba\begin{aligned}
A^{ab}\d g_{ab}=-\frac{\pd \math{L}}{\pd g^{ab}}\d g^{ab}\,,
\end{aligned}\ea
we can obtain
\ba\begin{aligned}
A_{ab}=-\frac{\pd \math{L}}{\pd g^{ab}}&=-2(E_R)^{cde}{}_{a} R_{cdeb}+(E_F)^{c}{}_{a}F_{bc}\\
&=-2(E_R)^{cde}{}_{(a} R_{|cde|b)}+(E_F)^{c}{}_{(a}F_{b)c}\,,
\end{aligned}\ea
which implies
\ba\begin{aligned}\label{antiscondition}
2(E_R)^{cde}{}_{[a} R_{|cde|b]}+(E_F)_{[a}{}^cF_{b]c}=0\,.
\end{aligned}\ea
Then, we have
\ba\begin{aligned}\label{ETheta}
\bm{E}_\f\d\f&=-\bm{\epsilon}\left(\frac{1}{2}T^{ab}\d g_{ab}+j^a\d A_a\right)
\end{aligned}\ea
with
\ba\begin{aligned}
T^{ab}&= 2 E_R^{cde(a}R_{cde}{}^{b)}+4\grad_c\grad_d E_R^{(a|c|b)d}-2E_F^{c(a}F^{b)}{}_c-g^{ab}\math{L}\,,\\
j^b&=2\grad_a E_F^{ab}.\\
\end{aligned}\ea
Next, we turn to calculate the expression of Noether current associated to the vector field $\z$,
\ba\begin{aligned}\label{Jexp}
\bm{J}_\z&=\bm{\Q}(\f, \math{L}_\z\f)-\z\.\bm{L}\\
&=\star (\bm{v}_\z-\z \math{L})\,,
\end{aligned}\ea
in which we have denoted $\bm{v}_\z=\left.\dbar \bm{v}\right|_{\d\f=\math{L}_\z\f}$. For the first term, we have
\ba\begin{aligned}
v_\z^c&=2E_R^{abcd}\grad_b(\math{L}_\z g_{ad})+2\math{L}_\z g_{bd} \grad_a E_R^{abcd}
+2 E_F^{cb}\math{L}_\z A_b\,.
\end{aligned}\ea
Using the expressions
\ba\begin{aligned}
\math{L}_\z g_{ab}=2\grad_{(a} \z_{b)}\,,\quad \math{L}_\z A_a=\grad_a(\z^b A_b)+\z^b F_{ba}\,,
\end{aligned}\ea
we have
\ba\begin{aligned}\label{vc}
v_\z^c&=4E_R^{abcd}\grad_b\grad_{(a}\z_{d)}+4\grad_{(b}\z_{d)} \grad_a E_R^{abcd}+2 E_F^{cb}\grad_b(\z^aA_a)+ 2E_F^{cb}\z^a F_{ab}\,.
\end{aligned}\ea
For the last two terms of above equation, we have
\ba\begin{aligned}
v_{34}^c=2 E_F^{cb}\grad_b(\z^aA_a)+ 2E_F^{cb}\z^a F_{ab}=2\grad_d(E_F^{cd}\z^aA_a)+j^c A_a\z^a+ 2E_F^{cb}\z^a F_{ab}.
\end{aligned}\ea
For the first two term of Eq. \eq{vc}, we have
\ba\begin{aligned}
v_{12}^c&=2E_R^{abcd}\grad_b\grad_{a}\z_{d}+2E_R^{abcd}\grad_b\grad_{d}\z_{a}
+2\grad_{b}\z_{d} \grad_a E_R^{abcd}+2\grad_{d}\z_{b} \grad_aE_R^{abcd}\\
&=2E_R^{abcd}\grad_{[b}\grad_{a]}\z_{d}+4E_R^{abcd}\grad_{[b}\grad_{d]}\z_{a}
+2E_R^{abcd}\grad_d\grad_b\z_{a}+2\grad_{b}\z_{d} \grad_a E_R^{abcd}+2\grad_{d}\z_{b} \grad_aE_R^{abcd}\\
&=E_R^{abdc}R_{abde}\z^e+2E_R^{abcd}R_{bdae}\z^{e}+2\grad_d(E_R^{abcd}\grad_b\z_{a})-2\grad_b\z_{a}\grad_dE_R^{abcd}
+2\grad_{b}\z_{d} \grad_a E_R^{abcd}+2\grad_{d}\z_{b} \grad_aE_R^{abcd}\\
&=E_R^{abdc}R_{abde}\z^e+2E_R^{abcd}R_{bdae}\z^{e}+2\grad_d(E_R^{abcd}\grad_b\z_{a})
-2\grad_b(\z_{a}\grad_dE_R^{abcd})+2\grad_{b}(\z_{d} \grad_a E_R^{abcd})+2\grad_{d}(\z_{b} \grad_aE_R^{abcd})\\
&\quad+2\z_{a}\grad_b\grad_dE_R^{abcd}-2\z_{d}\grad_{b}\grad_a E_R^{abcd}-2\z_{b}\grad_{d}\grad_aE_R^{abcd}\\
&=E_R^{abdc}R_{abde}\z^e+2E_R^{abcd}R_{bdae}\z^{e}+2\grad_d[E_R^{abcd}\grad_b\z_{a}+\z_{b} \grad_a(E_R^{abcd}-E_R^{dacb}-E_R^{bdca})]+2\z_d\grad_a\grad_b(E_R^{dacb}+ E_R^{abcd}-E_R^{bdca})\,.
\end{aligned}\ea
Using the fact that $E_R^{abcd}$ has the same symmetry as $R^{abcd}$ and therefore it also satisfies the Bianchi identity $E_R^{[abc]d}=0$, which implies
\ba\begin{aligned}
E_R^{abcd}+E_R^{bdca}+E_R^{dacb}=0\,,
\end{aligned}\ea
we have
\ba\begin{aligned}
v_{12}^c&=E_R^{abdc}R_{abde}\z^e+2E_R^{abcd}R_{bdae}\z^{e}+2\grad_d[E_R^{abcd}\grad_b\z_{a}+2\z_{b} \grad_aE_R^{abcd}]-4\z_d\grad_a\grad_bE_R^{bdca}\\
&=E_R^{abdc}R_{abde}\z^e+2E_R^{abcd}R_{bdae}\z^{e}+4\z_d\grad_{(a}\grad_{b)}E_R^{cadb}+4\z_d\grad_{[a}\grad_{b]}E_R^{cadb}
+2\grad_d[E_R^{abcd}\grad_b\z_{a}+2\z_{b} \grad_aE_R^{abcd}]\\
&=E_R^{abdc}R_{abde}\z^e+4\z_d\grad_{a}\grad_{b}E_R^{(c|a|d)b}+2E_R^{abcd}R_{bdae}\z^{e}
-2R_{abe}{}^cE_R^{eadb}\z_d-2R_{abe}{}^d E_R^{caeb}\z_d
+2\grad_d[E_R^{abcd}\grad_b\z_{a}+2\z_{b} \grad_aE_R^{abcd}].\\
\end{aligned}\ea
Using the relations
\ba\begin{aligned}
2E_R^{abcd}R_{bdae}\z^{e}&=E_R^{abcd}R_{bdae}\z^{e}+E_R^{abcd}R_{dabe}\z^{e}
=E_R^{abdc}R_{abde}\z^{e}\,,\\
2R_{abe}{}^cE_R^{eadb}\z_d&=R_{abe}{}^cE_R^{eadb}\z_d+R_{abe}{}^cE_R^{beda}\z_d=R_{abe}{}^cE_R^{abed}\z_d,\\
2R_{abe}{}^dE_R^{caeb}\z_d&=R_{abe}{}^dE_R^{caeb}\z_d+R_{eab}{}^dE_R^{caeb}\z_d=-R_{bea}{}^dE_R^{beac}\z_d,
\end{aligned}\ea
we have
\ba\begin{aligned}
v_{\z}^c&=3E_R^{abdc}R_{abde}\z^e-E_R^{abde}R_{abd}{}^c\z_e
+ 2E_F^{cb}\z^a F_{ab}+4\z_d\grad_{a}\grad_{b}E_R^{(c|a|d)b}
+j^c A_e\z^e\\
&\quad+2\grad_d(E_R^{abcd}\grad_b\z_{a}+2\z_{b} \grad_aE_R^{abcd}+E_F^{cd}\z^aA_a)\\
&=4E_R^{abd[c}R_{abd}{}^{e]}\z_e-2E_F^{b[c} F^{e]}{}_{b}\z_e+2E_R^{abd(c}R_{abd}{}^{e)}\z_e-2E_F^{b(c} F^{e)}{}_{b}\z_e+4\z_d\grad_{a}\grad_{b}E_R^{(c|a|d)b}\\
&\quad+j^c A_e\z^e+2\grad_d(E_R^{abcd}\grad_b\z_{a}+2\z_{b} \grad_aE_R^{abcd}+E_F^{cd}\z^aA_a)\\
&=2E_R^{abd(c}R_{abd}{}^{e)}\z_e-2E_F^{b(c} F^{e)}{}_{b}\z_e+4\z_d\grad_{a}\grad_{b}E_R^{(c|a|d)b}+j^c A_e\z^e+2\grad_d(E_R^{abcd}\grad_b\z_{a}+2\z_{b} \grad_aE_R^{abcd}+E_F^{cd}\z^aA_a)\,,\\
\end{aligned}\ea
where we have used Eq. \eq{antiscondition} in the last step. Combining the above results, we can further obtain
\ba\begin{aligned}
v_\z^d-\z^d\math{L}=\z^bT_b{}^d+\z^bA_b j^d-2\grad_a\left(E_F^{ad}A^c\z_c-2\grad_cE_R^{adbc}\z_b-E_R^{adbc}\grad_{[b}\z_{c]}\right)\,,
\end{aligned}\ea
which implies that
\ba\begin{aligned}
\bm{J}_\z=\bm{C}_\z+d\bm{Q}_\z\,,
\end{aligned}\ea
with
\ba\begin{aligned}
(\bm{C}_\z)_{a_2\cdots a_n}&=\bm{\epsilon}_{a a_2\cdots a_n}(\z^bT_b{}^a+\z^bA_b j^a)\,,\\
(\bm{Q}_\z)_{a_3\cdots a_n}&=\bm{\epsilon}_{aba_3\cdots a_n}\left(E_F^{ab}A^c\z_c-2\grad_dE_R^{abcd}\z_c-E_R^{abcd}\grad_{[c}\z_{d]}\right)\,.
\end{aligned}\nn\\\ea

\section{Integrability of the ADM mass in $\math{L}(g_{ab}, R_{abcd}, F_{ab})$ gravity}\label{secC}

In this section, we would like to show that
\ba\begin{aligned}
\int_{\inf}t\.\bm{\Q}(\f, \d\f)
\end{aligned}\ea
is integrable in the $n$-dimensional asymptotically flat spacetime of $\math{L}(g_{ab}, R_{abcd}, F_{ab})$ gravity when $\math{L}$ is an analytical function of the contraction of $R_{abcd}$ and $F_{ab}$.  Denote $y_i=R_{ab}^{cd}\cdots F_{ef}$ with $y_0=R$ being some quantity which is contracted by $R_{abcd}$ and $F_{ab}$. Then, we have
\ba\begin{aligned}
\math{L}=\math{L}(y_i)\,.
\end{aligned}\ea
For the asymptotically flat spacetimes, we have [Phys. Rev. D {\bf 50}, 846 (1994)]
\ba\begin{aligned}\label{sp1}
g_{ab}=\h_{ab}+\math{O}(1/r^{n-3})\,,\quad\quad \pd_a g_{bc}=\math{O}(1/r^{n-2})\,,\quad\quad A_a=\math{O}(1/r^{n-3})\,,\quad\quad
F_{ab}=\math{O}(1/r^{n-2})\,,
\end{aligned}\ea
which implies that
\ba\begin{aligned}\label{sp2}
\d g_{ab}=\math{O}(1/r^{n-3})\,,\quad\grad_a \d g_{cd}=\math{O}(1/r^{n-2})\,,\quad \d A_a=\math{O}(1/r^{n-3})\,.
\end{aligned}\ea
Then, we have $y_i=\math{O}(1/r^{n-2})$, and therefore
\ba\begin{aligned}
\math{L}_{(i)}(y_i)=\math{L}_{(i)}(0)+\math{O}(1/r^{n-2})\,.
\end{aligned}\ea
Using the Lagrangian, we have
\ba\begin{aligned}
E_R^{abcd}=\sum_{i}\math{L}_{(i)} E_i^{abcd}
\end{aligned}\ea
with
\ba\begin{aligned}
\math{L}_{(i)}=\frac{\pd \math{L}}{\pd y_i}\,,\quad\quad
E_i^{abcd}=\frac{\pd y_i}{\pd R_{abcd}}\,.
\end{aligned}\ea
Using the asymptotic conditions in Eqs. \eq{sp1} and \eq{sp2}, we have
\ba\begin{aligned}
E_i^{abcd}=\math{O}(1/r^{n-2})
\end{aligned}\ea
for $i>0$ in which $y_i$ contains at least one $F_{ab}$ or $R_{abcd}$. From this condition, it is not hard to see
\ba\begin{aligned}
r^{n-2}\left(\math{L}_{(i)}E_i^{abcd}\grad_b\d g_{ad}+\d g_{bd} \grad_a(\math{L}_{(i)} E_i^{abcd})\right)=0
\end{aligned}\ea
for $i>0$ at asymptotic infinity. Then we have
\ba\begin{aligned}
\int_{\inf}t\.\bm{\Q}^\text{grav}(\f, \d g)&=\int_{\inf}2t^e\bm{\epsilon}_{cea_3\cdots a_n}\left(\math{L}_0E_0^{abcd}\grad_b\d g_{ad}+\d g_{bd} \grad_a (\math{L}_0E_0^{abcd})\right)\\
&=2\math{L}_0(0)\int_{\inf}t^e\bm{\epsilon}_{cea_3\cdots a_n}\left(E_0^{abcd}\grad_b\d g_{ad}+\d g_{bd} \grad_a E_0^{abcd}\right)\\
&=2\math{L}_0(0)\int_{\inf}t^e\bm{\epsilon}_{cea_3\cdots a_n}E_0^{abcd}\grad_b\d g_{ad}\,,
\end{aligned}\ea
in which we used the expression
\ba\begin{aligned}
E_0^{abcd}=\frac{1}{4}(g^{ac}g^{bd}-g^{ad}g^{bc})\,.
\end{aligned}\ea
Then, with a same calculation of Eq. (86) by Iyer and Wald in [Phys. Rev. D {\bf 50}, 846 (1994)], this term can be expressed as
\ba\begin{aligned}
\int_{\inf}t\.\bm{\Q}^\text{grav}(\f, \d g)=-\d\left[\int_{\inf}dS[\pd_r g_{tt}-\pd_t g_{rt}+r^k h^{ij}(\pd_i h_{kj}-\pd_k h_{ij})]\math{L}_0(y_i)\right]\,.
\end{aligned}\ea
where $r^a=(\pd/\pd r)^a$ and $h_{ij}$ is the spatial metric. From Eqs. \eq{sp1} and \eq{sp2}, we can further obtain
\ba\begin{aligned}
E_F^{ab}=\math{O}(1/r^{n-2})\,,
\end{aligned}\ea
which implies that
\ba\begin{aligned}
\int_{\inf}t\.\bm{\Q}^\text{e.m.}(\f, \d A)=0\,.
\end{aligned}\ea
Combining the above results, we have
\ba\begin{aligned}
\int_{\inf}t\.\bm{\Q}(\f, \d\f)=\d\int_{\inf}t\.\bm{B}\,,
\end{aligned}\ea
in which $\bm{B}$ is chosen to be any $(n-1)$-form such that
\ba\begin{aligned}
t^a \bm{B}_{a a_3\cdots a_n}=-\math{L}_0(y_i)\tilde{\bm{\epsilon}}_{a_3\cdots a_n}[\pd_r g_{tt}-\pd_t g_{rt}+r^k h^{ij}(\pd_i h_{kj}-\pd_k h_{ij})]
\end{aligned}\ea
holds at asymptotical infinity. Here $\tilde{\bm{\epsilon}}$ is the induced volume element of the sphere at infinity.

\section{First law of black holes in $\math{L}(g_{ab}, R_{abcd}, F_{ab})$ gravity}\label{secD}

To derive the first law of black holes, we consider an asymptotically flat stationary black hole with Killing horizon $H$ which contains a bifurcate surface $\math{B}$. The Killing vector of $H$ is given by
\ba\begin{aligned}
\x^a=t^a+\W_H^{(\m)} \j^a_{(\m)}\,.
\end{aligned}\ea
Considering a variation from a stationary black hole to another stationary black hole, replacing $\zeta^a$ by the Killing vector field $\x^a$ and fixing $\x^a$, Eq. (22) of the letter reduces to
\ba\begin{aligned}\label{varid2}
d[\d \bm{Q}_\x-\x\.\bm{\Q}(\f,\d \f)]=0\,.
\end{aligned}\ea

As an asymptotically flat condition of the black hole, we can choose a gauge of the electromagnetic field such that $A_a\x^a$ vanishes at infinity. Under this gauge choice, the mass $M$, angular momentum $J_{(\m)}$, electric charge $Q$, and entropy $S$ are defined by
\ba\begin{aligned}\label{MJ}
M&=\int_\infty \left(\bm{Q}_t-t\.\bm{B}\right)\,,\quad J_{(\m)}=-\int_\inf  \bm{Q}_{\j_{(\m)}}\,,\\
Q&=-\int_{\inf} \bm{\epsilon}_{aba_3\cdots a_n} E_F^{ab}\,,
S=-2\p\int_\math{B}\bm{\epsilon}_{aba_3\cdots a_n} \bm{\hat{\epsilon}}_{cd} E_R^{abcd}\,,
\end{aligned}\ea
in which $\hat{\bm{\epsilon}}$ is the binormal of the bifurcate surface $\math{B}$

We  choose $\S$ as a hypersurface connecting a sphere at infinity and the bifurcate surface $\math{B}$. Then, integration of Eq. \eq{varid2} on $\S$ gives
\ba\begin{aligned}\label{first1}
\d M-\W_H\d J=\int_\math{B} \left[\d \bm{Q}_\x-\x\.\bm{\Q}(\f,\d \f)\right]\,.
\end{aligned}\ea
Then we have
\ba\begin{aligned}\label{first2}
\d M-\W_H\d J=&\d\int_\math{B}\bm{\epsilon}_{aba_3\cdots a_n}\left[E_F^{ab}A^c\x_c-2\x_c \grad_d E_R^{abcd}-E_R^{abcd}\grad_{[c}\x_{d]}\right]\\
&-\int_\math{B}\x\.\left[\bm{\Theta}^\text{e.m}(\f,\d\bm{A})+\bm{\Theta}^\text{grav}(\f,\d g)\right] \,.
\end{aligned}\ea

Assuming that $F_{ab}$, $g_{ab}$ and $\d g_{ab}$ are finite outside the singularity, we have $\bm{\Q}^\text{grav}(\f, \d g)$ and $\grad_d E_R^{abcd}$ being finite on the bifurcate surface $\math{B}$ because both of them are constructed by $F_{ab}$ and $g_{ab}$. However, since we choose the gauge such that $\x^a A_a$ vanishes at infinity, $A_a$ will not be finite at $\math{B}$. Considering the above results and the fact that $\x^a$ vanishes on $\math{B}$, Eq. \eq{first2} reduce to
\ba\begin{aligned}\label{first3}
\d M-\W_H\d J=&\d\int_\math{B}\bm{\epsilon}_{aba_3\cdots a_n}\left[E_F^{ab}A^c\x_c-E_R^{abcd}\grad_{[c}\x_{d]}\right]-\int_\math{B}\x\.\bm{\Theta}^\text{e.m}(\f,\d\bm{A}) \,.
\end{aligned}\ea
For the first term in the right side of Eq. \eq{first3}, we have
\ba\begin{aligned}
\int_\math{B} \bm{\epsilon}_{aba_3\cdots a_n}\left[E_F^{ab}A^c\x_c-E_R^{abcd}\grad_{[c}\x_{d]}\right]
&=-\F_H \int_\math{B} \bm{\epsilon}_{aba_3\cdots a_n} E_F^{ab}-\int_\math{B} \bm{\epsilon}_{aba_3\cdots a_n}E_R^{abcd}\grad_{[c}\x_{d]}\\
&=\F_H Q-\int_\math{B} \bm{\epsilon}_{aba_3\cdots a_n}E_R^{abcd}\grad_{[c}\x_{d]}\,.
\end{aligned}\ea
Performing a calculation similar to [Phys. Rev. D {\bf 50}, 846 (1994)], we have
\ba\begin{aligned}
\d \int_\math{B} \bm{Q}_\x=\d(\F_H Q)+\frac{\k}{2\p} \d S\,.
\end{aligned}\ea
For the second term in the right side of Eq. \eq{first3}, we have
\ba\begin{aligned}
\int_\math{B} \x\.\bm{\Q}^\text{e.m.}(\f,\d \bm{A})&=2\int_\math{B}\x^c\bm{\epsilon}_{aca_3\cdots a_n}E_F^{ab}\d A_b\\
&=2\int_\math{B} \tilde{\bm{\epsilon}}\x^c\hat{\bm{\epsilon}}_{ac}E_F^{ab}\d A_b,
\end{aligned}\ea
where $\tilde{\bm{\epsilon}}$ is the induced volume element of $\math{B}$.
To evaluate the above expression, we consider another cross-section $B'$ of the horizon and finally make $B'\to \math{B}$ and define $s^a$ as another null normal vector field on cross-section of $H$ which satisfies
\ba\begin{aligned}
s^a \x_a=-1,\quad\quad s_as^a=0\,,\quad\quad \math{L}_\x s^a=0\,.
\end{aligned}\ea
Then, we have $\hat{\bm{\epsilon}}=s\wedge \x$ on $B'$, which gives
\ba\begin{aligned}\label{cef}
\x^c\hat{\bm{\epsilon}}_{ac}E_F^{ab}\d A_b&=-\x^cs_c\x_aE_F^{ab}\d A_b\\
&=\x_aE_F^{ab}\d A_b\,.
\end{aligned}\ea
Assume $k^a$ and $l^a$ are two finite vector fields proportional to $\x^a$ and $s^a$ on $H$, respectively, i.e., we have
\ba\begin{aligned}
k^a l_a=&-1,\quad\quad k_ak^a=0,\quad\quad l_a l^a=0\,,\\
&k^a=C\x^a,\quad\quad l^a=C^{-1} s^a\,,
\end{aligned}\ea
where $C$ is a scalar field on $H$ and goes to infinity on $\math{B}$ since $\x^a$ vanishes on $\math{B}$. We denote $w^a$ as a finite vector fields tangent to $B'$ and assume $\math{L}_\x w^a=0$ on $H$. Then, we have
\ba\begin{aligned}\label{EFxz}
\x_a w_b E_F^{ab}=C^{-1}k_a w_b E_F^{ab}\,.
\end{aligned}\ea
Noting that $E_F^{ab}$, $k_a$ and $w^a$ are all finite outside the singularity, the right side of Eq. \eq{EFxz} will vanish when $B'$ approaches $B$. Considering that the right side of Eq. \eq{EFxz} is constant on $H$, we must have $\x_a z^i_b E_F^{ab}=0$ on $H$. Thus, $\x_aE_F^{ab}s_b=E_F^{ab}k_a l^b$ is the only non-vanishing component of $\x_aE_F^{ab}$. Then, Eq. \eq{cef} gives
\ba\begin{aligned}
\x_aE_F^{ab}\d A_b&=-\x_aE_F^{ab}s_b\x^c\d A_c\\
&=\x_a s_b E_F^{ab}\d \F_H\\
&=-\frac{1}{2}\hat{\bm{\epsilon}}_{ab} E_F^{ab}\d\F_H\,.
\end{aligned}\ea
Therefore, we have
\ba\begin{aligned}
\int_\math{B} \x\.\bm{\Q}(\f,\d \f)&=-\d\F_H \int_\math{B}\tilde{\bm{\epsilon}}\hat{\bm{\epsilon}}_{ab} E_F^{ab}\\
&=-\d\F_H \int_\math{B}\bm{\epsilon}_{aba_3\cdots a_n} E_F^{ab}\\
&=Q \d\F_H\,.
\end{aligned}\ea
Combining the above results, we can get
\ba\begin{aligned}
\int_\math{B} \left[\d \bm{Q}_\x-\x\.\bm{\Q}(\f,\d \f)\right]=\F_H\d Q+\frac{\k}{2\p}\d S\,.
\end{aligned}\ea
Substituting the above result to Eq. \eq{first3}, we can finally obtain the first law
\ba\begin{aligned}
\d M=T \d S+\W_H^{(\m)}\d J_{(\m)}+\F_H \d Q\,.
\end{aligned}\ea

\section{WCCC in quadratic gravity}\label{SecE}

The Lagrangian density of the quadratic gravity coupled to the Maxwell field is given by
\ba\begin{aligned}
\math{L}=\frac{1}{2} R+c_1 R^2+c_2 R_{ab}R^{ab}+c_3 R_{abcd}R^{abcd}-\frac{1}{4}F_{ab}F^{ab}\,.
\end{aligned}\ea
The source-free equations of motion are given by
\ba\begin{aligned}\label{Hab}
&2 E_R^{cde(a}R_{cde}{}^{b)}+4\grad_c\grad_d E_R^{(a|c|b)d}-2E_F^{c(a}F^{b)}{}_c-g^{ab}\math{L}=0\,,
&2\grad_a E_F^{ab}=0\,.
\end{aligned}\ea
We set the spherically symmetric static line element and gauge field as
\ba\begin{aligned}\label{ds2app}
ds^2=-f(r)dv^2+2\m(r) dv dr+r^2(d\q^2+\sin^2\q d\f^2)\,,\quad\quad \bm{A}=\Y(r) dv\,.
\end{aligned}\ea
Here we choose the ingoing Eddington coordinates in the equation of motion $\grad_aF^{ab}=0$, it is easy to get
\ba\begin{aligned}\label{Aapp}
A_a=-\left(\int_{\inf}^{r}\frac{q \m(r)}{r^2}\text{dr}\right)(dv)_a\,,
\end{aligned}\ea
where $q$ is an integral constant related to the electric charge of the spacetime.

\subsection{Central singularity of the spherically symmetric static black holes}\label{SecE1}

To discuss the violation of WCCC, let's first prove that there is a spacetime singularity at the center $r=0$ of the spherically symmetric static charged black holes in quadratic gravity. Using the line element \eq{ds2app}, it is not hard to get
\ba\begin{aligned}\label{R2}
R_{abcd}R^{abcd}&=\frac{4}{r^4}+\frac{f''(r)^2}{\mu (r)^4}-\frac{8 f(r) f'(r) \mu '(r)}{r^2 \mu (r)^5}+\frac{4 f'(r)^2}{r^2 \mu (r)^4}+\frac{f'(r)^2 \mu '(r)^2}{\mu (r)^6}\\
&-\frac{2 f'(r) f''(r) \mu '(r)}{\mu (r)^5}+\frac{4 f(r)^2}{r^4 \mu (r)^4}-\frac{8 f(r)}{r^4 \mu (r)^2}+\frac{8 f(r)^2 \mu '(r)^2}{r^2 \mu (r)^6}\,.
\end{aligned}\ea
Lacking of a spacetime singularity at $r=0$ demands that $R_{abcd}R^{abcd}$ is a finite quantity on $r=0$. We assume that the leading terms of $f(r)$ and $\m(r)$ at $r\to0$ are given by
\ba\begin{aligned}
f(r)\propto f_{-n_1}r^{-n_1}\,,\quad\quad \m(r)\propto \m_{-n_2}r^{-n_2}\,,
\end{aligned}\ea
in which $n_1, n_2$ are some integers. Then, the leading term of $R_{abcd}R^{abcd}$ can be further obtained, and it is given by
\ba\begin{aligned}
R_{abcd}R^{abcd}\propto \frac{4}{r^4}+\frac{A_1 f_{-n_1}^2 r^{-4-2(n_1-2n_2)}}{\m_{-n_2}^4}-\frac{8f_{-n_1} r^{-4-(n_1-2n_2)}}{\m_{-n_2}^2}
\end{aligned}\ea
with
\ba\begin{aligned}
A_1=\left(n_1 \left(n_1+2\right)+5\right) n_1^2-2 \left(n_1^2+n_1+4\right) n_2 n_1+\left(n_1^2+8\right) n_2^2+4\,.
\end{aligned}\ea
For the case without central singularity, we need that $R_{abcd}R^{abcd}$ is finite when $r\to0$, which implies that we must have
\ba\begin{aligned}
n_1=2n_2\,.
\end{aligned}\ea
Then, we have
\ba\begin{aligned}
R_{abcd}R^{abcd}\propto \frac{\left(1-y\right)^2+\left(\left(n_2+1\right){}^2+2\right) n_2^2 y^2}{r^4}
\end{aligned}\ea
with
\ba\begin{aligned}
y=\frac{f_{-2n_2}}{\m_{-n_2}^2}\,.
\end{aligned}\ea
The regular condition that $R_{abcd}R^{abcd}$ is finite at $r=0$ implies that $n_1=n_2=0$ and $f_0=\mu_0^2\neq0$. The value of $f_0$ and $\m_0$ can be arbitrarily small but they cannot strictly be  zero since if $f_0=\m_0=0$, the leading term will become $n_1, n_2\geq 1$ and they cannot cancel out the divergent term $4/r^4$. Therefore, for the regular case, we can expand
\ba\begin{aligned}
f(r)=\sum_{n=0}^{\inf} f_n r^n\,,\quad\quad \m(r)=\sum_{n=0}^{\inf} \m_n r^n
\end{aligned}\ea
at $r=0$. Using the above results, the equation of motion $\left.r^4H^{vr}\right|_{r=0}=0$ gives $q=0$. That is to say, for the case with $q\neq 0$, there must be a spacetime singularity at $r=0$.

\subsection{Near-horizon solution of charged static extremal black hole}\label{SecE2}
Now, we turn to evaluate the near-horizon solution of the charged static extremal black hole. Assuming that $f(r)$ and $\m(r)$ are analytical functions at $r=r_h$, we can expand $f(r), \m(r)$ at $r=r_h$ as
\ba\begin{aligned}
f(r)=\sum_{n=0}^{\inf} \frac{f^{(n)}(r_h)}{n!}(r-r_h)^n\,,\quad\quad \m(r)=\sum_{n=0}^{\inf} \frac{\m^{(n)}(r_h)}{n!}(r-r_h)^n\,.
\end{aligned}\ea
For the extremal black hole solution, we have
\ba\begin{aligned}
f(r_h)=f'(r_h)=0\,.
\end{aligned}\ea
Using the above expression, it is not hard to obtain
\ba\begin{aligned}
R_{abcd}R^{abcd}=\frac{4}{r_h^2}+\frac{f''(r_h)}{\m(r_h)}\,.
\end{aligned}\ea
at $r=r_h$. Assuming that the geometry is regular at horizon, we have $\m(r_h)$ being finite. Performing the coordinate transformation $v\to v/\m(r_h)$, we can set $\m(r_h)=1$. Using the equations of motion $H^{vr}(r_h)=0$ and $\pd_r H^{vr}(r_h)=0$, we can further obtain
\ba\begin{aligned}\label{f2q}
f''(r_h)=\frac{2}{r_h^2}\,,\quad\quad r_h=\frac{q}{\sqrt{2}}\,.
\end{aligned}\ea
Using the above results and together with the equation of motion $\pd_r H^{\q\q}(r_h)=0$, we can get
\ba\begin{aligned}
f^{(3)}(r_h)=\frac{6\r'(r_h)-12}{r_h^3}\,.
\end{aligned}\ea
Substituting them into the equation of motion
\ba\begin{aligned}
\frac{4(2c_1+c_2+2c_3)r_h^4}{8(7c_1+2c_2+c_3)-r_h^2}\pd_r^2 H^{\q\q}(r_h)+H^{vv}(r_h)=0\,,
\end{aligned}\ea
we can obtain
\ba\begin{aligned}
\m'(r_h)=-\frac{8(c_2+4c_3)}{r_h(11c_2+48c_3+r_h^2)}\,.
\end{aligned}\ea
Therefore, we have
\ba\begin{aligned}
f^{(3)}(r_h)=-\frac{12(16c_2+64c_3+r_h^2)}{r_h^3(12c_2+48c_3+r_h^2)}\,.
\end{aligned}\ea
Repeating the above processes, the higher-order terms $f^{n}(r_h)$ and $\m^{(n)}(r_h)$ can also be obtained. However, since these higher-order terms don't contribution to $S'_\text{ext}(r_h)$, we are not going to solve for them here.

\subsection{Constraint from WCCC}\label{SecE3}

Finally, we perform the above results to discuss the distruction condition $S'(r_h)< 0$. To obtain $S'(r_h)$, we need to evaluate $S(r)$ for any sphere $\math{S}$ with radius $r$. Using the expression of the Wald entropy, we can further obtain
\ba\begin{aligned}
S(r)=-2\p\int_\math{S} \bm{\epsilon}_{aba_1a_2}\hat{\bm{\epsilon}}_{cd}E_R^{abcd}=-32\p^2 r^2 \m(r)^2 E_R^{vrvr}(r)\,.
\end{aligned}\ea
Together with the near-horizon solution obtained in the above, we can further obtain
\ba\begin{aligned}
S'_\text{ext}(r_h)=\frac{16\p^2}{r_h}[r_h^2+4(c_2+4c_3)]\,.
\end{aligned}\ea

Since there is a central singularity of the charged spacetime solutions, the WCCC demands that all black holes in this theory cannot be destroyed. From the destruction condition $S'_\text{ext}(r)<0$, we can see that if $c_2+4c_3<0$, the extremal black holes with $r_h<-4(c_2+c_3)$ will violate the WCCC. The validity of WCCC demands that
\ba\begin{aligned}
c_2+4c_3\geq 0\,.
\end{aligned}\ea

\section{WCCC in a cubic gravity}\label{SecF}
We consider a four-dimensional cubic gravity coupled to the Maxwell field with the Lagrangian density
\ba\begin{aligned}\label{LGcubic}
\math{L}=\frac{1}{2} R+c_1 R_{ab}^{cd}R_{cd}^{ef}R_{ef}^{ab}+c_2 R R_{ab}^{cd}R_{cd}^{ab}-\frac{1}{4}F_{ab}F^{ab}\,.
\end{aligned}\ea
The source-free equations of motion are also given by $H_{ab}=0$ and $\grad_a F^{ab}=0$ in which $H_{ab}$ is shown in Eq. \eq{Hab}.

\subsection{Central singularity of the spherically symmetric static black holes}\label{SecF1}

First, we would like to prove that there is a spacetime singularity at the center $r=0$ of the spherically symmetric static charged black holes in cubic gravity with the Lagrangian \eq{LGcubic}. From the discussions in Sec. \ref{SecE1}, the condition $R_{abcd}R^{abcd}$ is finite at $r=0$ demands that
\ba\begin{aligned}\label{expfm}
f(r)=\sum_{n=0}^{\inf} f_n r^n\,,\quad\quad \m(r)=\sum_{n=0}^{\inf} \m_n r^n
\end{aligned}\ea
with $f_0=\m_0^2\neq0$ at $r=0$. For the regular case, we also need to demand $R$ and $R_{ab}^{cd}R_{cd}^{ef}R_{ef}^{ab}$ are finite at $r=0$. Using Eq. \eq{expfm}, it is not hard to obtain
\ba\begin{aligned}
&R\propto \frac{8\m_1-6 f_1}{r}\,,
\end{aligned}\ea
which implies that we must have $f_1=(4/3)\m_1$ for the regular case. Using this result, we can also obtain
\ba\begin{aligned}
R_{ab}^{cd}R_{cd}^{ef}R_{ef}^{ab}\propto -\frac{16\m_1^3}{9r^3}+\cdots\,,
\end{aligned}\ea
which gives $f_1=\m_1=0$ for the regular case. Substituting the results $f_0=\m_0^2\neq0, f_1=\m_1=0$ into Eq. \eq{expfm}, the equation of motion $\left.r^6 H^{rv}\right|_{r=0}=0$ demands $q=0$. That is to say, for the case $q\neq0$, there must be a singularity at $r=0$.

\subsection{Near-horizon solution of charged static extremal black hole and constraint from WCCC}\label{SecF2}

With a similar calculation of quadratic gravity in Sec. \ref{SecE2}, it is not hard to obtain the extremal spherically symmetric static solution of this theory,
\ba\begin{aligned}\label{sol2}
q^2&=\frac{2(r_h^4+24c_1+16c_2)}{r_h^2}\,,\quad\quad f(r_h)=f'(r_h)=0\,,\quad\quad f''(r_h)=\frac{2}{r_h^2}\,,\\
&f^{(3)}(r_h)=\frac{B_1}{r_h^3C}\,,\quad\quad \m(r_h)=1\,,\quad\quad \m'(r_h)=\frac{B_2}{r_h C}\,,
\end{aligned}\ea
with
\ba\begin{aligned}
B_1&=12 [8 \left(97 c_1+30 c_2\right) r_h^{12}-512 \left(c_1+2 c_2\right) \left(873 c_1^2+3024 c_2 c_1+2108 c_2^2\right) r_h^4\\
&\quad -64 \left(435 c_1^2+1456 c_2 c_1+1140 c_2^2\right) r_h^8-49152 \left(c_1+2 c_2\right){}^2 \left(99 c_1^2+429 c_2 c_1-10 c_2^2\right)-r_h^{16}]\,,\\
B_2&=48 [\left(25 c_1+14 c_2\right) r_h^{12}-8 \left(141 c_1^2+440 c_2 c_1+284 c_2^2\right) r_h^8-192 \left(c_1+2 c_2\right) \left(75 c_1^2+272 c_2 c_1+180 c_2^2\right) r_h^4\\
&\quad -1536 \left(c_1+2 c_2\right)^2 \left(153 c_1^2+648 c_2 c_1+76 c_2^2\right)]\,,\\
C&=\left(24 c_1+48 c_2+r_h^4\right){}^2 \left(64 \left(c_2-3 c_1\right) r_h^4+64 \left(63 c_1^2+228 c_2 c_1+268 c_2^2\right)+r_h^8\right)\,.\\
\end{aligned}\ea
Using this near-horizon solution, we can further obtain
\ba\begin{aligned}
S'_\text{ext}(r_h)=\frac{8\p^2}{r_h^3}\frac{r_h^8-96 c_1 r_h^4+192 \left(9 c_1^2+28 c_2 c_1+20 c_2^2\right)}{24 \left(c_1+2 c_2\right)+r_h^4}\,.
\end{aligned}\ea
From the expression of $q^2$ in Eq. \eq{sol2}, we can see that $x>0\,\,\text{and}\,\, x>-8(3c_1+2c_2)$ with $x=r_h^4$ for the extremal black hole solutions. Since there is a central singularity for the charged spacetime solutions in this theory, the validity of WCCC demands that $S'(r_h)\geq 0$ for all possible extremal black hole, which implies that we should have
\ba\label{SSS}
\frac{(x-x_1)(x-x_2)}{x-x_0}\geq0
\ea
for any $x>0 \,\,\text{and}\,\, x\geq-8(3c_1+2c_2)$. Here we have denoted
\ba\begin{aligned}
x_0=-24 \left(c_1+2 c_2\right)\,,\quad x_1=8\left(6c_1-\sqrt{3(c_1-10c_2)(3c_1+2c_2)}\right)\,,\quad x_2=8\left(6c_1+\sqrt{3(c_1-10c_2)(3c_1+2c_2)}\right).
\end{aligned}\ea
Next, we analyze this in the following categories:\\
(1) If $3c_1+2c_2>0\,\,\text{and}\,\, (c_1-10c_2)<0$, we need $x_0\leq 0$, which gives
\ba\begin{aligned}
c_2>0\,\,\text{and}\,\,-\frac{2c_2}{3}<c_1<10 c_2\,.
\end{aligned}\ea
(2) If $3c_1+2c_2>0\,\,\text{and}\,\, (c_1-10c_2)\geq0$, we know that $c_1>0$ and therefore $x_2>0$, i.e., there exists some $x$ which does not satisfy Eq. \eq{SSS}.\\
(3) If $3c_1+2c_2<0\,\,\text{and}\,\, (c_1-10c_2)>0$, which gives $c_2<0$, we need
\ba\label{SSSS}
\frac{(x-x_1)(x-x_2)}{x-x_0}\geq0
\ea
for any $x\geq z=-8(3c_1+2c_2)$. It is easy to verify that $x_0-z=-32c_2>0$, i.e., $x_0>z$, which implies that there exists some $x$ which does not satisfy Eq. \eq{SSSS}.\\
(4) If $3c_1+2c_2<0\,\,\text{and}\,\, (c_1-10c_2)\leq0$, we need
\ba\label{SSSS}
\frac{(x-x_1)(x-x_2)}{x-x_0}\geq0
\ea
for any $x\geq z=-8(3c_1+2c_2)$. Therefore, we need $x_0<z\,\,\text{and}\,\, x_2<z$, which gives
\ba\begin{aligned}
c_2>0\,\,\text{and}\,\, c_1<-\frac{2c_2}{3}.
\end{aligned}\ea
(4) If $3c_1+2c_2=0$, we need
\ba\label{SSSS}
x-48c_1>0
\ea
for any $x>0$, which implies that $c_1\leq0$. Therefore, we have
\ba\begin{aligned}
c_2\geq 0\,\,\text{and}\,\,  c_1=-\frac{2c_2}{3}\,.
\end{aligned}\ea
Combining the above results, we finally obtain
\ba\begin{aligned}
c_2\geq0\,\,\text{and}\,\, c_1\leq 10c_2\,.
\end{aligned}\ea

\end{document}